 \documentclass[pmlr,twocolumn,10pt]{jmlr} 

\usepackage{booktabs}
\usepackage{siunitx}

\usepackage[switch]{lineno}

\newcommand{\equal}[1]{{\hypersetup{linkcolor=black}\thanks{#1}}}

\theorembodyfont{\upshape}
\theoremheaderfont{\scshape}
\theorempostheader{:}
\theoremsep{\newline}

\title[]{
Robust AI-ECG for Predicting  Left Ventricular Systolic Dysfunction in Pediatric Congenital Heart Disease
}
\author{%
\Name{Yuting Yang} \Email{Yuting.Yang@childrens.harvard.edu}\\
\addr Computational Health Informatics Program, Boston Children’s Hospital, Department of Pediatrics, Harvard
Medical School, Boston, MA, USA
\AND
\Name{Lorenzo Peracchio} \Email{lorenzo.peracchio01@universitadipavia.it}\\
\addr Department of Electrical, Computer and Biomedical Engineering, University of Pavia, Pavia, Italy
\AND
\Name{Joshua Mayourian} \Email{Joshua.Mayourian@childrens.harvard.edu}\\
\addr Department of Cardiology, Boston Children’s Hospital, Department of Pediatrics, Harvard Medical School,
Boston, MA, USA
\AND
\Name{John K. Triedman} \Email{John.Triedman@cardio.chboston.org}\\
\addr Department of Cardiology, Boston Children’s Hospital, Department of Pediatrics, Harvard Medical School,
Boston, MA, USA
\AND
\Name{Timothy Miller}\equal{Co-senior}\Email{Timothy.Miller@childrens.harvard.edu}\\
\addr Computational Health Informatics Program, Boston Children’s Hospital, Department of Pediatrics, Harvard
Medical School, Boston, MA, USA
\AND
\Name{William G. La Cava}\footnotemark[1] \Email{William.LaCava@childrens.harvard.edu}\\
\addr Computational Health Informatics Program, Boston Children’s Hospital, Department of Pediatrics, Harvard
Medical School, Boston, MA, USA
}

\begin{document}

\maketitle

\begin{abstract}
Artificial intelligence-enhanced electrocardiogram (AI-ECG) has shown promise as an inexpensive, ubiquitous, and non-invasive screening tool to detect left ventricular systolic dysfunction in pediatric congenital heart disease. However, current approaches rely heavily on large-scale labeled datasets, which poses a major obstacle to the democratization of AI in hospitals where only limited pediatric ECG data are available. In this work, we propose a robust training framework to improve AI-ECG performance under low-resource conditions. Specifically, we introduce an on-manifold adversarial perturbation strategy for pediatric ECGs to generate synthetic noise samples that better reflect real-world signal variations. Building on this, we develop an uncertainty-aware adversarial training algorithm that is architecture-agnostic and enhances model robustness. Evaluation on the real-world pediatric dataset demonstrates that our method enables low-cost and reliable detection of left ventricular systolic dysfunction, highlighting its potential for deployment in resource-limited clinical settings.

\end{abstract} 

\section{Introduction}
\label{sec:intro}
Congenital heart disease (CHD) refers to structural or functional heart abnormalities present at birth. It is one of the most common birth defects, affecting approximately 1\% of live births worldwide \citep{van2011birth}. Electrocardiogram (ECG) is a rapid, standardized, and cost-effective tool widely used for diagnosing cardiovascular diseases and initial cardiac screening \citep{saarel2018electrocardiograms}. Existing studies have shown that Artificial intelligence-enhanced electrocardiogram (AI-ECG) can reliably detect early markers of cardiovascular dysfunction, including left ventricular systolic dysfunction (LVSD) in the general adult population \citep{attia2019screening,naser2024artificial}, which is commonly associated with heart failure and adverse cardiovascular outcomes. 

However, AI-ECG applications in pediatric cardiology remain largely unexplored. Pediatric ECGs differ significantly from adult ECGs in both epidemiology and characteristics, which may limit the generalizability of adult AI-ECG models \citep{siontis2021detection}. Existing work in pediatric congenital heart disease \citep{mayourian2024cmr,mayourian2025lvef} requires large amounts of labeled training data. Privacy concerns and regulatory restrictions make data sharing challenging, and large-scale publicly available pediatric ECG datasets are lacking. Consequently, hospitals with limited ECG data face challenges in developing reliable, site-specific models, highlighting the need for models that are robust in data-scarce scenarios. 

\begin{figure*}[h]
\floatconts
  {fig:method}
  {\caption{The overall framework of the proposed approach. 
  It identifies ``Borderline ECGs", i.e., those near the classification boundary, augments them with on-manifold adversarial perturbations, and trains the model using a combination of original and adversarial samples to improve robustness.}}
  {\includegraphics[width=1\linewidth]{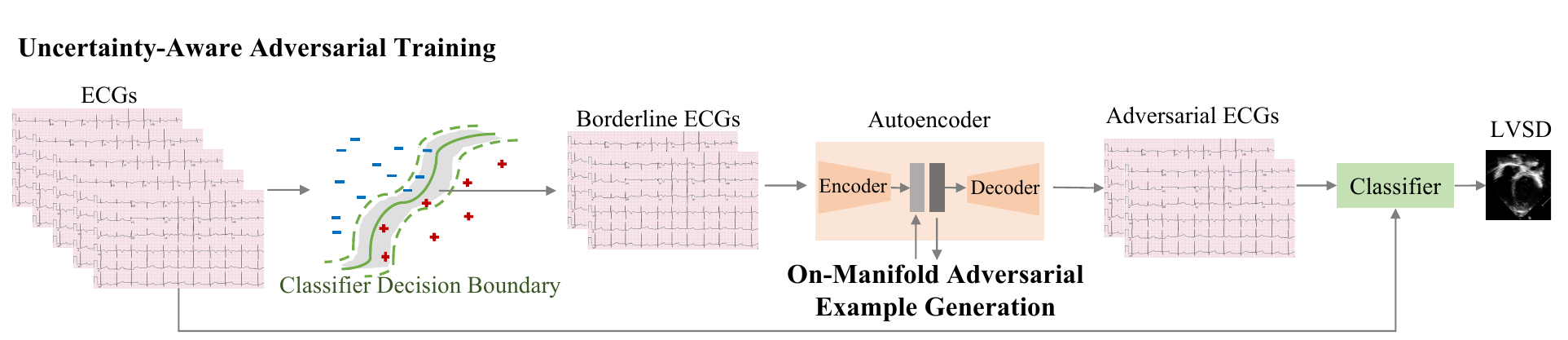}}
\end{figure*}

To address this challenge, this study proposes a robust AI-ECG approach, incorporating the principles of adversarial training as shown in Figure \ref{fig:method}. We design an adversarial training algorithm to finetune the existing AI-ECG model with generated adversarial perturbations on the model's most uncertain samples (those near the model's decision boundary). This uncertainty-aware adversarial training focuses the model’s learning on its most vulnerable regions. It enables the model to learn more robust and intrinsic features, thereby achieving better generalization even in low-sample scenarios. In addition, we propose an on-manifold adversarial example generation algorithm that generates perturbations constrained by the latent data manifold learned by an autoencoder \citep{he2022masked}. Compared to perturbations in the raw signal domain, embedding-space perturbations tend to remain closer to the manifold of physiologically plausible ECGs, leading to more realistic variations. Extensive experiments on real-world dataset demonstrate that our model exhibits enhanced robustness and can achieve competitive performance using only 10\% of the original dataset, particularly within some lesion subgroups (e.g., patients with pacemakers).

In summary, our contributions are the following:

\begin{enumerate}
\item We introduce on-manifold adversarial perturbation generation for pediatric ECGs, enabling the synthesis of noise samples that more closely resemble real-world signals. 
\item We propose an uncertainty-aware adversarial training algorithm, which is not limited to specific model architectures and can be used to enhance model robustness under limited data conditions.
\item Validation on real-world dataset shows that our method can achieve low-cost and reliable detection for left ventricular systolic dysfunction in pediatric patients.
\end{enumerate}

\section{Background and Related Work}
\label{sec:rw}
Despite advances in congenital heart surgery and medical treatment over the decades, CHD remains a leading cause of mortality in newborns \citep{drews2020spontaneous}. These conditions place a significant burden on patients and their families, highlighting CHD as a critical global public health issue.

Early diagnosis and management are crucial for improving outcomes and quality of life for affected individual. Given the limited evidence-based therapies for heart failure in congenital heart disease, developing preventive approaches through the accessible and low-cost detection of early markers, such as left ventricular systolic dysfunction (LVSD), is of significant interest. LVSD is defined as a condition characterized by left ventricular ejection fraction (LVEF) of less than 40\%, indicating impaired ventricular contraction and reduced blood ejection from the left ventricle during each heartbeat \citep{kemp2012pathophysiology}. LVSD is associated with a $>$8-fold increased risk of subsequent heart failure and nearly a 2-fold risk of premature death \citep{sangha2023detection}. Early identification of LVSD allows for timely interventions like medical therapy and potentially improving heart failure symptoms and mortality \citep{anjewierden2024detection}. 

Prior studies have demonstrated the potential of AI-ECG to screen for LVSD in the general adult population \citep{attia2019screening,naser2024artificial}. 
Recently, a few studies have begun to explore the application of AI-ECG methods for detecting ventricular dilation and dysfunction in pediatric congenital heart disease \citep{mayourian2024cmr,mayourian2025lvef}. \citet{mayourian2025lvef} trained a convolutional neural network on more than 120,000 ECGs to detect left ventricular ejection fraction (LVEF) of 40\% and achieve high model performance. However, these models rely on large-scale labeled datasets to reach high efficiency and generalizability. This requirement creates a significant barrier for many small hospitals, where pediatric ECG data is often scarce, making it difficult to develop reliable diagnostic models.

Although some studies have begun to enlarge ECG datasets with data augmentation \citep{xu2022hygeia,nonaka2021randecg}, they usually exploited some general augmentation techniques such as adding Gaussian noise, which are typically developed and validated on adults. \citet{wiedeman2024decorrelative} applied adversarial training to improve the model's resistance to adversarial perturbations for adults. Such approaches fail to account for the unique physiological and pathological characteristics of pediatric CHD patients. Overall, the robustness of ECG-based diagnostic models in low-resource pediatric settings remains insufficiently studied.

\section{Methods}
\label{sec:me}

Let $X$ denote the set of training ECGs and $Y$ their corresponding labels. The training dataset can be written as $\mathcal{D} = \{(x, y)| x\in X, y\in \mathcal{Y}\}$. The objective of an AI-ECG model is to learn a predictive model $f_\theta:X \rightarrow \mathcal{Y}$, parameterized by $\theta$, which maps ECG inputs to task-specific outputs.

\subsection{On-Manifold Adversarial Example Generation}
\label{sec:me:apg}

The purpose of adversarial example generation is to perturb a normal input $x$ to generate an adversarial example $x_{adv}=x+\delta$ for a target model (e.g., a LVSD detector), so that $x_{adv}$ preserves the semantic of $x$ while misleading the target model $f_\theta$ into making incorrect predictions:
\begin{equation}
    f_\theta(x + \delta) \neq f_\theta(x)
\end{equation}

The loss function of generating adversarial examples ($\mathcal{L}_{adv}$) is:
\begin{equation}\label{eq:ladv}
    \mathcal{L}_{adv}(x,y,\delta)=\ell(f_\theta(x + \delta),y)-\lambda*d(x+\delta,x)
\end{equation}
where $\ell$ is cross-entropy loss function and 
$d$ is the regularizer to constrain the perturbation $\delta$ to not change the original semantic of $x$ after adding perturbation. $\lambda$ is used to balance these two losses. It was set to 0.1 by default. $d$ can be cosine similarity function between original ECG signal $x$ and perturbed signal $x+\delta$:
\begin{equation}
    d(x+\delta,x)=\frac{\langle x, x+\delta \rangle}{\|x\|_2 \, \|x+\delta\|_2}
\end{equation}

Thus, the optimization objective of adversarial example generation is:
\begin{equation}\label{eq:loss_adv}
    \max_{\delta} \mathcal{L}_{adv}(x,y,\delta)
\end{equation}

We exploit Projected Gradient Descent (PGD, \citet{PGD}) to optimize $\delta$ in \equationref{eq:loss_adv}, which iteratively maximizes the loss function based on the gradients of the input $\nabla_{\delta} \mathcal{L}_{adv}$. 
Let $\operatorname{Clip}_{\varepsilon}$ define a projection back to the infinity norm ball by clamping $\delta$ to $\varepsilon$. Thus, 
\begin{equation}
    \delta^{t+1}=\operatorname{Clip}_{\varepsilon}(\delta^t+\alpha \cdot \text{sign}(\nabla_{\delta} \mathcal{L}_{adv}(x,y,\delta^t)))
\end{equation}
where $\alpha$ is the learning rate. After $T$ steps, we get the optimal perturbation $\delta^T$, which can mislead the original prediction of $x$. 

As in Smooth Adversarial Perturbations \citep{adv_ecg}, we employ convolution to smooth the generated signal, which takes the weighted average of one position of the signal and its neighbors. We take the adversarial perturbation as the parameter and add it to the clean examples after convolving with a number of Gaussian kernels. We denote $G(s,\sigma)$  to be a Gaussian kernel with size $s$ and standard deviation $\sigma$. The resulting adversarial example can be written as a function of $\delta$:

\begin{equation}
    x_{\mathrm{adv}}=x+\frac{1}{M} \sum_{m=1}^M \delta^T \circledast G(s[i], \sigma[i])
\end{equation}
$M$ is the number of Gaussian kernels. 

Perturbing raw ECG signals often produces physiologically implausible waveforms. In contrast, embedding-space perturbations preserve the ECG’s semantic and structural properties, as the encoder maps signals onto a smooth manifold and the decoder constrains them within the distribution of plausible patterns. To obtain latent representations of ECGs, we pre-train an autoencoder (ViT-MAE, \citet{he2022masked}) for pediatric ECGs. Instead of perturbing the raw signal $x$ directly \citep{adv_ecg}, we encode each ECG sequence into a continuous latent representation $z = Enc(x)$ and then apply perturbations in $z$. 
Thus, the objective in \equationref{eq:ladv} becomes:
\begin{align}\label{loss_adv_emb}
    \mathcal{L}_{adv}(x,y,\delta) =\ell(f_\theta(Dec(Enc(x) + \delta)),y)\nonumber\\
    -\lambda*d(Dec(Enc(x)+\delta),x)
\end{align}
where $Dec$ decodes a latent representation into the original input space. Then same optimization process with PGD is applied to optimize $\mathcal{L}_{adv}$ in Equation \ref{loss_adv_emb}. 

\subsection{Uncertainty-Aware Adversarial Training}\label{sec:uaat}
Adversarial training was designed to improve the model's resistance to adversarial perturbations. \citet{adv_training} proposed a min-max optimization training algorithm which formalizes robustness enhancement problem as a saddle point problem:
\begin{equation}\label{eq:1}
	\min_{\theta}[\mathbb{E}_{(x,y)\sim \mathcal{D}}\max_{\delta} \mathcal{L}_{adv}(x, y,\delta)]
\end{equation}
This is an \emph{inner maximization} problem and an \emph{outer minimization} problem. The inner maximization problem is a process of generating adversarial examples, aiming to find a perturbation $\delta$ that fools the victim model most or achieves a high training loss. The outer minimization problem then learns to improve the ability of predicting under the perturbations. The goal of the outer minimization problem is to find model parameters so that the ``adversarial loss'' given by the inner problem is minimized. When the parameters $\theta$ yield a (nearly) vanishing risk, the corresponding model is perfectly robust to attacks. The inner maximization objective is optimized by the adversarial perturbation generation algorithm in Section \ref{sec:me:apg}.

Since deep models exhibit varying vulnerabilities across different regions, we propose an uncertainty-aware adversarial training strategy. At each training iteration, we estimate the model’s uncertainty for all training samples and retain only those with high uncertainty. Adversarial perturbations are then generated on these uncertain samples to encourage the model to attend to vulnerable regions and to learn smoother decision boundaries. We quantify the model’s predictive uncertainty using the entropy of its output distribution. For an input $x$, let $p_\theta(y|x)$ denote the predicted probability over $C$ classes.
The uncertainty $\mathcal{U}(x)$ is computed as:
\begin{equation}
    \mathcal{U}(x) = - \sum_{c=1}^C p_\theta(y=c|x) \log p_\theta(y=c|x)
\end{equation}
Then the training samples $\mathcal{D}_{u}$ for each iteration is:
\begin{equation}
    \mathcal{D}_u = \mathbf{TopK({x\in \mathcal{D},\mathcal{U}(x),k})}
\end{equation}
Then, the final training loss function $\mathcal{L}$ is:
\begin{equation}
    \mathcal{L} = \mathbb{E}_{(x,y)\sim \mathcal{D}_u}\max_{\delta} \mathcal{L}_{adv}(x, y,\delta)
\end{equation}
As the model's parameters is optimized in each iteration, the most uncertain samples and its corresponding adversarial examples are also different during the training process. This allows the training process to continually explore additional vulnerable regions of the model, forcing the learned decision boundary to become smoother and ultimately leading the model to capture more intrinsic and robust features.

\section{Experiments}
\label{sec:ex}

\subsection{Dataset}
We used patient data from a large children's hospital in the United States, collected up to January 2023. 
The patient inclusion criterion was the availability of at least one echocardiogram with a recorded LVEF. To enrich the training set with overlapping pathophysiology relevant to pediatric heart failure and to broaden applicability across the heterogeneous patient population encountered in pediatric cardiology, we also included patients with cardiomyopathy as well as patients without congenital heart disease. 

\begin{table}[t]
\floatconts
  {tab:data_stat}
  {\caption{Characteristics of training and testing cohorts (n (\%)).}}
  {%
  \scriptsize
  \begin{tabular}{p{3cm}rr}
  \toprule
  & Training & Testing \\
  \midrule
  \textbf{ECGs} & & \\
  Totals & 124265 & 54230 \\
  Tetralogy of fallot & 8980 (7.2\%) & 4108 (7.6\%) \\
  Cardiomyopathy & 18509 (14.9\%) & 8082 (14.9\%) \\
  Atrial septal defect & 14860 (12.0\%) & 6171 (11.4\%) \\
  Complete atrioventricular canal & 2640 (2.1\%) & 1130 (2.1\%) \\
  Coarctation of the aorta & 12353 (9.9\%) & 4998 (9.2\%) \\
  Double outlet right ventricular & 2720 (2.2\%) & 1042 (1.9\%) \\
  D-loop TGA & 4313 (3.5\%) & 1982 (3.7\%) \\
  Ebstein & 1481 (1.2\%) & 676 (1.2\%) \\
  Hypoplastic left heart syndrome & 4050 (3.3\%) & 1689 (3.1\%) \\
  L-loop TGA & 1191 (1.0\%) & 399 (0.7\%) \\
  Pulmonary atresia & 3994 (3.2\%) & 1808 (3.3\%) \\
  Total anomalous pulmonary venous return & 1463 (1.2\%) & 758 (1.4\%) \\
  Tricuspid atresia & 1117 (0.9\%) & 358 (0.7\%) \\
  Truncus arteriosus & 1461 (1.2\%) & 548 (1.0\%) \\
  Ventricular septal defect & 21171 (17.0\%) & 8926 (16.5\%) \\
  Dextrocardia & 1265 (1.0\%) & 403 (0.7\%) \\
  Pacemaker & 2732 (2.2\%) & 1145 (2.1\%) \\
  \midrule
  \textbf{Patient-level Characteristics} & & \\
  Patients & 49158 & 21068 \\
  Male & 26311 (53.5\%) & 11251 (53.4\%) \\
  Female & 22835 (46.5\%) & 9813 (46.6\%) \\
  Missing & 12 ($<0.1\%$) & 4 ($<0.1\%$) \\
  \midrule
  \textbf{Outcomes} & & \\
  LVEF $\leq$ 50\% & 8525 (6.9\%) & 3674 (6.8\%) \\
  LVEF $\leq$ 40\% & 3381 (2.7\%) & 1473 (2.7\%) \\
  LVEF $\leq$ 30\% & 1490 (1.2\%) & 598 (1.1\%) \\
  \bottomrule
  \end{tabular}
  }
\end{table}

All raw ECG signals were retrieved from the MUSE ECG data management system (GE Healthcare, Chicago, IL, USA). CHD lesions were identified according to the institutional Fyler coding system. Paced patients were identified based on ECG diagnoses of dual-chamber or ventricular pacing. ECG recordings shorter than 10 seconds or missing lead information were excluded. Fewer than 2\% of ECGs failed quality control, typically due to random issues such as unintentionally disconnected leads. The remaining ECGs were resampled to 250 Hz, high-pass filtered, and truncated to 2048 samples (approximately 8 seconds) to facilitate use with convolutional neural networks. Each ECG has 12 leads. Additional details of quality control and preprocessing have been described previously \citep{mayourian2024pediatric}. The training cohort comprised 124,265 ECGs (49,158 patients; median age 10.5 years
(IQR 3.5–16.8); 46.5\% patients were female and 53.5\% were male). The testing cohort comprised 54,230 ECGs (21,068 patients; median age 10.9 years (IQR 3.7–17.0); 46.6\% patients were female and 53.4\% were male). 24.1\% patients had congenital heart disease in the overall testing cohort. The characteristics of the dataset are summarized in Table \ref{tab:data_stat}.

\paragraph{Outcomes} 
LVEF values were extracted from echocardiogram reports, with the left ventricle consistently corresponding to the morphological left ventricle. 
LVEF was determined using the bullet method \citep{odellAccuracyLeftVentricular2019}. 
The primary outcome was LVEF of 40\% or less (quantitatively at least moderate dysfunction). Secondary outcomes included LVEF of 50\% or less (quantitatively at least mild dysfunction) and LVEF of 30\% or less (quantitatively severe dysfunction). The reports also provide outcomes of LVEF $\leq$ 45\% and LVEF $\leq$ 35\%. The median LVEF was 62.0\% (IQR 57.4\%–66.0\%) in training set and 62.0\% (IQR 57.6\%–66.0\%) in test set, where 2.7\% ECGs had an LVEF of 40\% or less.

\subsection{Baselines}
\paragraph{ResNet} 
We use the state-of-the-art AI-ECG model for pediatric LVSD detection as the baseline model \citep{mayourian2024pediatric}. The model is based on ResNet, a convolutional neural network originally designed for image recognition, which can process ECG signals by capturing hierarchical temporal features through residual connections. More specifically, the ResNet consisted of a convolutional layer followed by 4 residual blocks with 2 convolutional layers per block. The convolutional layers start with 64 filters for the first layer and residual block, with a filter increase and subsampling. The output of each convolutional layer is rescaled using batch normalization and fed into a rectified linear activation unit, with subsequent dropout at a rate of 0.2. Max pooling and convolutional layers with filter length 1 are included in the skip connections to match main branch signal dimensions. The output of the last block is fed into a fully connected layer with a sigmoid activation function given that outcomes are not mutually exclusive.
\paragraph{ResNet+DA} We further construct an augmented baseline by applying data augmentation (DA) to ResNet. Specifically, for each training sample, we construct an augmented sample by introducing Gaussian noise that stimulates the real-life noise. We add random Gaussian noise at four real-world ECG noise frequency ranges: 3-12 Hz (motion artifact during tremors), 12-50 Hz (lower-frequency muscle activation artifact), 50-100 Hz (electrode motion noise), and 100-150 Hz (higher-frequency muscle activation artifact) \citep{dhingra2025artificial,khunte2023detection}. Powerline interference noise is further captured within the 50-100 Hz and 100-150 Hz frequency ranges \citep{friesen1990comparison}. Then ResNet is trained with both original and augmented samples.

\subsection{Metrics}
Due to data privacy concerns, it is challenging to obtain existing pediatric ECG datasets from smaller hospitals to validate the effectiveness of our approach. As an alternative, we compared the performance of the prediction model when trained on the full dataset versus a smaller subset consisting of our data. To emulate real-world scenarios in hospitals of varying sizes, we randomly selected 10\% of the original training set to form smaller training subsets. We apply adversarial training described in Section \ref{sec:uaat} while training and test on the whole test set. Given the imbalanced dataset, both area under the receiver operating curve (AUROC, \citet{fawcett2006introduction}) and area under the precision-recall curve (AUPRC, \citet{saito2015precision}) are computed to evaluate the model's performance. 

\subsection{Implementation Details}
While training, we use Adam optimizer. A maximum of 100 epochs was used with early stopping on the basis of validation loss. Final hyperparameters were kernel size 17, batch size 64, and learning rate 0.001. For adversarial training, we keep Top K=30\% uncertain samples for each training iteration. Inner optimization steps $T$ is 20, $\alpha$=0.001 and clamping bound $\delta$ is 0.5. Following \citet{adv_ecg}, $s$ is set to [5, 7, 11, 15, 19] and $\sigma$ is [1, 3, 5, 7, 10]. 

\begin{figure*}[h]
\floatconts
  {fig:overall}
  {\caption{Model performance of ResNet, ResNet+DA (data augmentation), and ResNet+ADV (adversarial training) on overall and pacemaker cohorts under full and 10\% training data.}}
  {\includegraphics[width=1\linewidth]{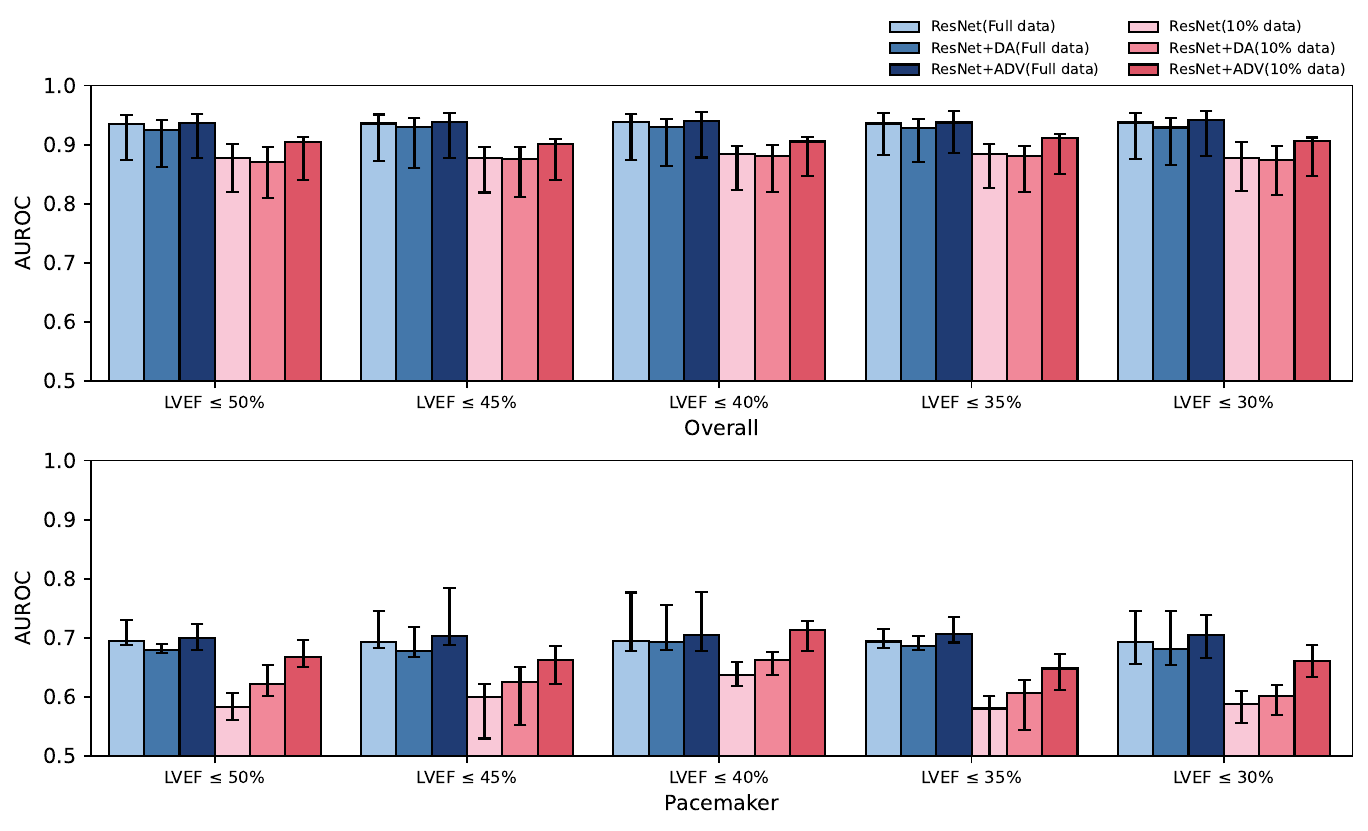}}
\end{figure*}
To provide image-based inputs suitable for the ViT-MAE, ECGs were transformed into spectrograms. Specifically, each of the 12 ECG leads was converted using the Short-Time Fourier Transform (STFT, \citet{huang2019ecg}), retaining both the real and imaginary components of the resulting spectra. This process yielded 24 channels per ECG recording, which were treated as the input image channels for the model. Since the pretrained base ViT-MAE was originally designed for three-channel RGB images, we adapted it to handle 24-channel spectrograms by replicating the weights of the first convolutional projection layer across the additional channels, while leaving the remainder of the encoder–decoder architecture unchanged. The model was then adapted to the ECG domain via self-supervised continual learning on the spectrograms, using the mean squared error loss of the reconstructions. This strategy leveraged the pretrained backbone as a strong initialization, while enabling the model to progressively refine its latent representations for ECGs.

\subsection{Evaluation Results}

There are 18 lesion subgroups in both training and testing cohorts. In structural lesions, such as Coarctation of the aorta or Ventricular septal defect, ECG abnormalities (e.g., chamber hypertrophy, axis deviation, repolarization changes) more directly reflect the underlying hemodynamic burden. These lesion-specific signatures allow AI-ECG to capture physiologically meaningful features associated with impaired ventricular function, potentially yielding more consistent performance. In contrast, for patients with pacemakers, the ECG is dominated by pacing artifacts and non-physiologic ventricular activation, which may obscure native conduction and repolarization patterns linked to ventricular function. As a result, LVEF prediction in this cohort represents a greater challenge, but also serves as an important proof-of-concept for the robustness and generalizability of AI-ECG. Thus, we present model performances on overall cohort as well as pacemaker subgroup in Figure \ref{fig:overall}. More evaluations across all lesions can be found in Appendix \ref{apd:all_lesions}.

Figure \ref{fig:overall} compares three models, ResNet, ResNet+DA and ResNet+ADV (ours), across varying training sizes (full and 10\% data) in LVEF outcomes with different thresholds. The performance of ``ResNet(Full data)'' represents a performance upper-bound benchmark for the task, such large-scale data collections are rarely available in most hospitals. Confidence intervals (CIs) were obtained via resampling with 1000 bootstraps. 95\% CIs are shown using bootstrapping and indicated by error bars.

We find that models trained on the full dataset (``ResNet(Full data)'') consistently outperform those trained on only 10\% of the data (``ResNet(10\% data)''). For the overall cohort, performance decreased from AUROC = 0.94(0.87-0.95) to 0.88(0.82-0.89) in predicting LVEF $\leq$ 40\%, corresponding to an absolute drop of 0.06 median AUROC. The decline was larger for pacemaker lesion with LVEF $\leq$ 35\% or LVEF $\leq$ 30\% , where AUROC decreased by about 0.11 ($\simeq$16\%), subgroups further complicated by highly scare positive samples. Specifically, patients with pacemakers account for only 2.1\% of the overall cohort, representing a more data-scarce scenario. In addition, compared with LVEF $\leq$ 40\%, patients reaching LVEF $\leq$ 35\% or $\leq$ 30\% are even fewer, as these thresholds reflect more severe left ventricular dysfunction. These findings highlight current AI-ECG model's sensitivity to data scarcity and suggest potential limitations in resource-constrained clinical settings.

We then evaluate the effect of adversarial training (``ResNet+ADV''). Under ideal conditions with abundant training data (``Full data''), adversarial training achieves comparable or slightly improved performance relative to the baseline models for both overall or pacemaker cohort. In contrast, under data-scarce settings (e.g., using only 10\% of the training data), adversarial training yields more substantial performance gains. With adversarial training, ResNet can achieve comparable performance with that trained with full data: the difference of ``ResNet(Full data)'' and ``ResNet+ADV(10\%)'' is within a margin of 0.03 median AUROC across all outcomes for overall cohort.

The improvement is particularly significant for the highly underrepresented pacemaker lesion. For example, in predicting LVEF $\leq$ 30\%, ``ResNet+ADV(10\% data)'' achieves an AUROC of 0.67 (0.64-0.70), significantly outperforming ``ResNet(10\% data)'' (AUROC=0.59 (0.56-0.61)) by 0.08 and nearly matching ``ResNet(Full data)'' (AUROC=0.69 (0.65-0.74) with only a 0.02 difference in the median. Given that pacemaker ECGs are dominated by pacing-induced patterns rather than native conduction, these results indicate that adversarial training can enhance model robustness specifically in lesion groups with atypical ECG characteristics. Overall, these findings demonstrate that our approach maintains comparable or slightly improved performance under data-rich conditions, while yielding substantial gains under data-scarce conditions.

\begin{table*}[hbtp]
\centering
\caption{Comparison of training strategies on pacemaker subgroup. Variants share the same training framework but differ in one component: ``w/o uncertainty'' generates adversarial examples for all samples; ``w/o on-manifold'' applies perturbations directly to input ECG signals.}
\label{tab:abl_study}
\resizebox{\textwidth}{!}{ 
\begin{tabular}{l|rr|rr|rr} 
\toprule
 & \multicolumn{2}{c|}{LVEF $\leq$ 50\%} 
 & \multicolumn{2}{c|}{LVEF $\leq$ 40\%} 
 & \multicolumn{2}{c}{LVEF $\leq$ 30\%} \\
 & \multicolumn{1}{c}{AUROC} & \multicolumn{1}{c|}{AUPRC} & \multicolumn{1}{c}{AUROC} & \multicolumn{1}{c|}{AUPRC} & \multicolumn{1}{c}{AUROC} & \multicolumn{1}{c}{AUPRC} \\
\midrule
ResNet & 0.58 (0.56–0.61) & 0.23 (0.09-0.40) & 0.64 (0.62–0.66) & 0.22 (0.09-0.41) & 0.59 (0.56–0.61) & 0.19 (0.07-0.39) \\
\midrule
ResNet+ADV & \textbf{0.68 (0.66–0.71)} & \textbf{0.29 (0.14-0.47)} & \textbf{0.73 (0.69–0.74)} & \textbf{0.28 (0.13-0.46)} & \textbf{0.67 (0.64–0.70)} & \textbf{0.26 (0.09-0.46)} \\
\ \ w/o uncertainty & 0.65 (0.64–0.68) & 0.26 (0.11-0.44) & 0.69 (0.67–0.69) & 0.23 (0.09-0.44) & 0.65 (0.61–0.67) & 0.25 (0.06-0.44) \\
\ \ w/o on-manifold & 0.65 (0.63–0.69) & 0.26 (0.11-0.45) & 0.68 (0.66–0.69) & 0.24 (0.10-0.46) & 0.63 (0.62–0.65) & 0.22 (0.06-0.43) \\
\bottomrule
\end{tabular}
} 
\end{table*}

\subsection{Ablation Studies} We perform ablation experiments to evaluate the impact of different components in our approach. As shown in Table \ref{tab:abl_study}, compared with ResNet+ADV, ``w/o uncertainty'' generates adversarial examples for all inputs without leveraging uncertainty $\mathcal{U}(x)$ to select borderline samples. 
``w/o on-manifold'' uses the same adversarial example generation algorithm but generates perturbations directly on the input ECG signals, rather than the latent space learned via autoencoder.

All ablation experiments show a decrease in AUROC and AUPRC compared with the full adversarial model (``ResNet+ADV''), indicating that each component contributes to enhancing the model’s robustness. Meanwhile, these ablated models still outperform the baseline without adversarial training (``ResNet''), further demonstrating the effectiveness and robustness of our proposed adversarial training framework. Moreover, the impact of individual modules on the final performance varies: replacing the on-manifold perturbation module with perturbations applied directly in the ECG signal space produces the largest performance drop, particularly in data-scarce scenarios such as LVEF $\leq$ 30\%. This further underscores the importance of generating realistic augmentations to improve model robustness.


\subsection{Model Explainability}

\begin{table}[hbtp]
\centering
\caption{Distributional discrepancies between testing cohorts and three training cohorts: original training set (``Org''), adversarially perturbed training set (``Adv''), and a mixed dataset combining both original and adversarial samples (``Combined'').}
\label{tab:interpretability}
\resizebox{\columnwidth}{!}{ 
\begin{tabular}{l|ccc|cc} 
\toprule
& \multicolumn{3}{c|}{Global Discrepancy} & \multicolumn{2}{c}{Local Discrepancy} \\
& Center(pos) & Center(neg) & MMD & JSD & KLD  \\
\midrule
Org & 0.6257 & 0.5623 & 0.0038 & 0.3950 & 8.0197\\
Adv & \textbf{0.5721} & \textbf{0.4964} & \textbf{0.0034} & 0.4020 & 8.1413\\
Combined & 0.5934 & 0.5293 & 0.0035 & \textbf{0.3290} & \textbf{6.1033}\\
\bottomrule
\end{tabular}
} 
\end{table}

To better understand why adversarial training enhances model robustness under data-scarcity scenarios, we analyzed the distributional discrepancies between the training and testing data. Intuitively, a lower  discrepancy between training and testing distributions facilitates better generalization of the model. To this end, we measured the discrepancy between the testing data and three training scenarios: the original training set (``Org'' in Table \ref{tab:interpretability}), the adversarially perturbed training set (``Adv''), and a mixed dataset combining both original and adversarial samples (``Combined''). We employed two categories of discrepancy measures to quantify both global and local relationships between datasets in the latent space. For global similarity, we use ``Center(pos)'' and ``Center(neg)'', which measure the distance between the centroids of positive and negative class samples, respectively, as well as the Maximum Mean Discrepancy (MMD, \citet{gretton2012kernel}). MMD evaluates the difference between two distributions by comparing the mean embeddings of samples in a reproducing kernel Hilbert space, and has been widely used for distribution alignment. For local discrepancy, we exploit the Jensen–Shannon Divergence (JSD, \citet{lin2002divergence}) and Kullback–Leibler Divergence (KLD, \citet{kullback1951information}). JSD symmetrizes and smooths KL divergence to quantify the overlap between two probability distributions, while KLD measures the relative entropy, i.e., how one distribution diverges from another.

As shown in Table \ref{tab:interpretability}, adversarial training markedly reduces the global distributional gap between the training and test sets, as indicated by lower center distances and MMD values. However, it simultaneously introduces small bias at the local distribution level, reflected by marginally higher JSD and KLD. In contrast, combining the original and adversarial data balances these effects: the mixture substantially improves local similarity while maintaining global alignment. This hybrid strategy leads to richer and more diverse representations, resulting in better overall alignment with the test distribution. These findings suggest that while adversarial training is effective in capturing global structures, integrating it with original data is crucial to alleviating local distributional bias and achieving more robust generalization.


\section{Conclusion and Future Work}
\label{sec:dis}
In this work, we propose a robust training framework to improve AI-ECG robustness in low-resource settings. Our approach combines an on-manifold adversarial perturbation strategy for pediatric ECGs with an uncertainty-aware adversarial training algorithm, which identifies borderline samples near the classification boundary and augments them to enhance model robustness. Evaluation on real-world pediatric datasets demonstrates reliable and cost-effective detection of left ventricular systolic dysfunction, highlighting its potential for deployment in resource-limited clinical environments.

As our proposed method represents an effective training framework that is not limited to specific architectures or tasks, it can be readily applied to other healthcare scenarios. For example, it could be extended to Echocardiography (Echo) for assessing cardiac function, or to more complex, multi-modal settings that integrate ECG, Echo, and other cardiac-related data. In future work, we plan to extend this approach to a broader range of clinical tasks to further evaluate its generalization capability. Our goal is to establish it as a generalizable and convenient tool that can be adopted by different institutions, enabling the deployment of site-specific, effective AI models and promoting the democratization of AI in healthcare. 

\acks{
Research reported in this publication was supported by the National Library Of Medicine of the National Institutes of Health under Award Number R01LM012973 and R01LM014300. This work was partially supported by the Kostin Innovation Fund at Boston Children's Hospital. The content is solely the responsibility of the authors and does not necessarily represent the official views of the National Institutes of Health.
}

\bibliography{jmlr-sample}

\appendix

\section{Evaluation across all lesions}\label{apd:all_lesions}
\begin{figure}[h]
\floatconts
  {fig:lesions}
  {\caption{Model performance across congenital heart disease lesion subgroups (VSD = Ventricular septal defect, COA=Coarctation of the aorta, HLHS = Hypoplastic left heart syndrome, CAVC = Complete atrioventricular canal, DORV = Double outlet right ventricular, TAPVR = Total anomalous pulmonary venous return).}}
  {\includegraphics[width=1\columnwidth]{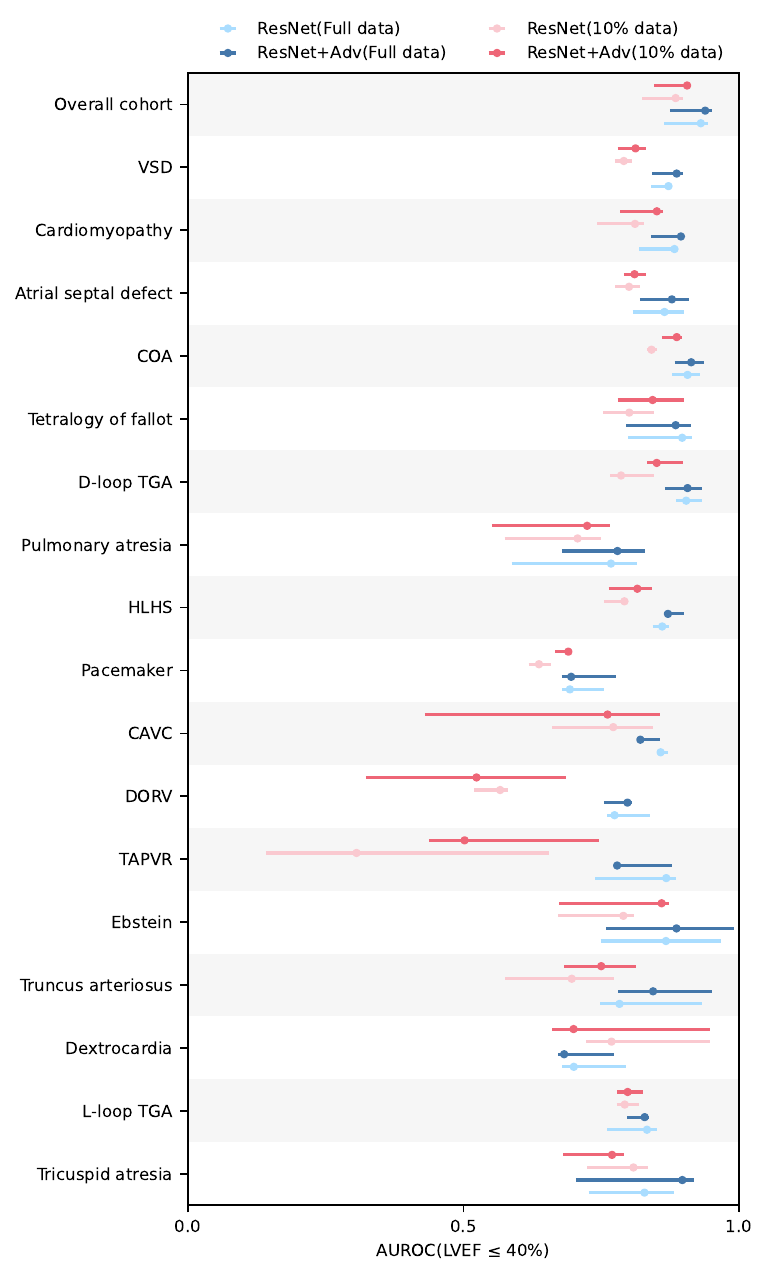}}
\end{figure}
\end{document}